\documentstyle[11pt,paspconf,epsf]{article}

\begin{document}

\title{New VLBI Constraints for 0957+561 Lens Models}

\author{Deborah Haarsma}
\affil{Calvin College, Grand Rapids, MI 49546  (dhaarsma@calvin.edu)}
\author{Joseph Leh\'ar}
\affil{Harvard-Smithsonian Center for Astrophysics, Cambridge, MA 02138}
\author{Rennan Barkana}
\affil{Institute for Advanced Study, Princeton, NJ 08540}

\begin{abstract}
Time-delay measurements of 0957+561 based on radio monitoring (Haarsma
et al.\ 1999) now agree with optical monitoring (Oscoz et al.\ 1997;
Kundic et al.\ 1997).  Recent models (Barkana et al.\ 1999; Bernstein \&
Fischer 1999) incorporate many recent observations, but the systematic
uncertainties in the models still dominate the uncertainty in the
cosmological results. VLBI observations of the
milli-arcsec jet structure have provided the most important set of
modeling constraints.  We present new observations of this structure,
made at 18 cm with the Very Long Baseline Array, the Very Large Array,
and the Green Bank 140 ft.\ telescope.  Compared to the data set of
Garrett et~al.\ (1994), ours has a similar theoretical noise level
(20$\mu$Jy), but nearly twice as many baselines (66 compared to 36) and
more uniform coverage of the UV plane.  Our initial maps confirm the basic
jet structures seen by Garrett et al., with comparable RMS noise
(60-80$\mu$Jy), and are in reasonable agreement with the jet structure
model of Barkana et al.\ (1999).  The maps also hint at a new compact
component leaving the core, and a previously undetected diffuse
component at the end of the jet.  Further refinements in
fringe-fitting, mapping, and self-calibration should yield
significantly lower RMS noise (closer to the theoretical level).  We
will use a modified form of the CalTech VLBI Modelfit program (written
by Barkana et al.\ 1999) to determine the magnification matrix between
the images.  The refined magnification matrix should provide stronger
constraints on the lens model, and thus reduce the uncertainty
in the cosmological results.
\end{abstract}



\begin{figure}
\plotfiddle{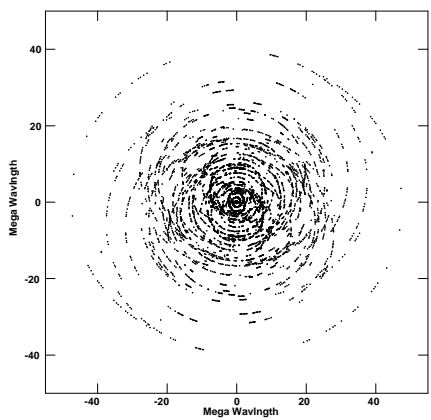}{2in}{0}{30}{30}{-110}{-40}
\caption{UV coverage (66 baselines), $\lambda=18$~cm.}
\end{figure}

\begin{figure}
\plottwo{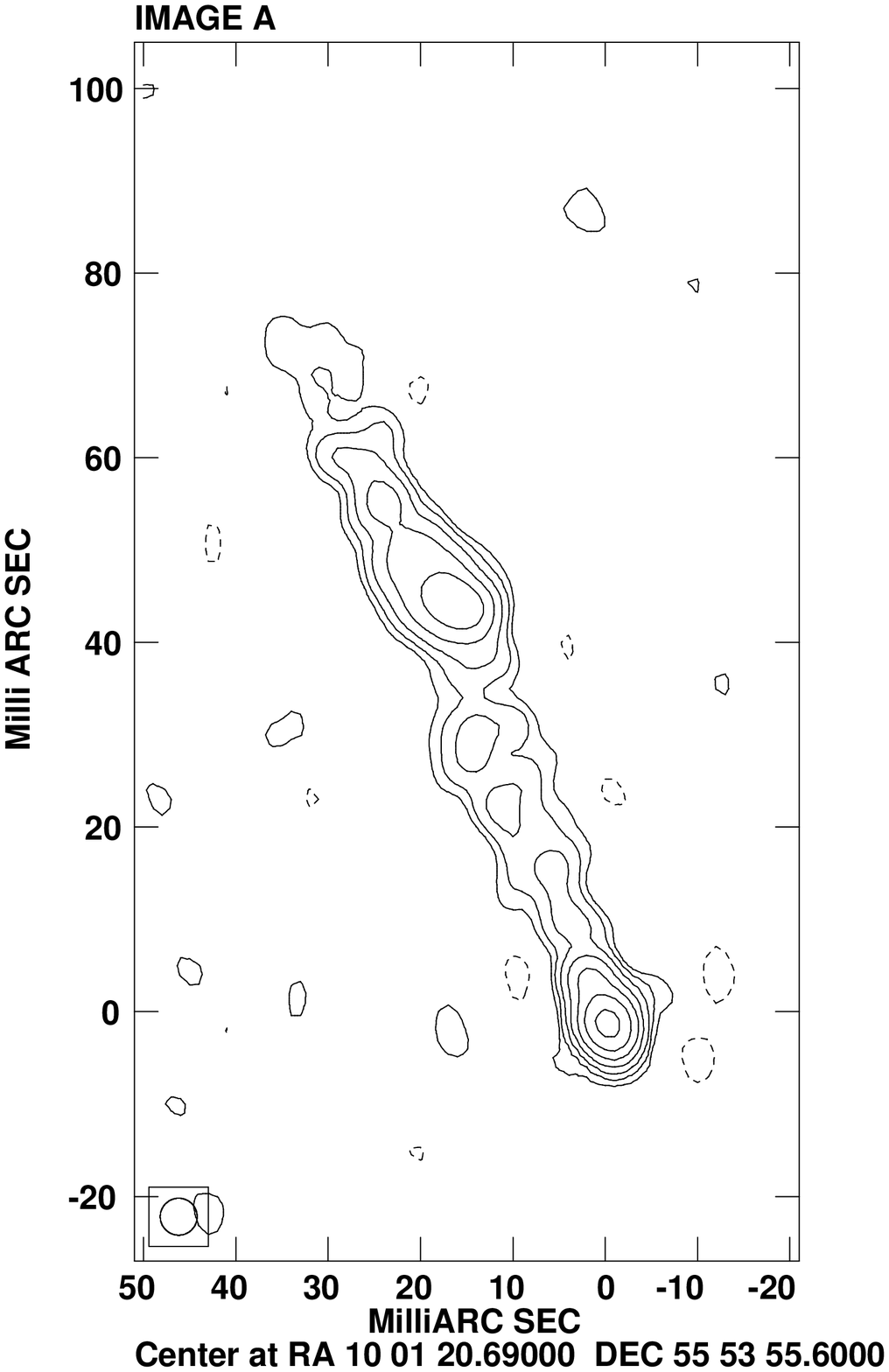}{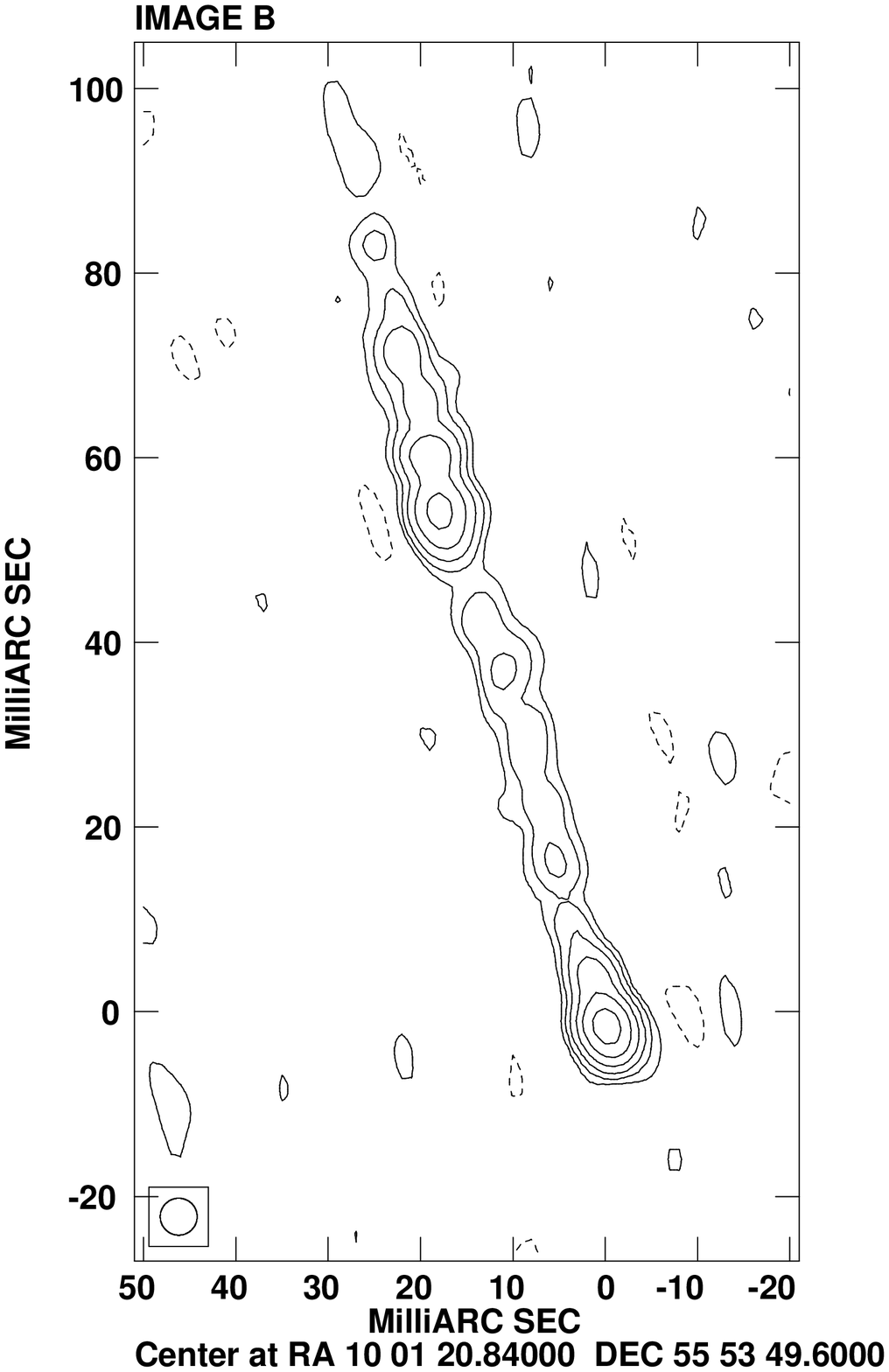}
\caption{Maps of 0957+561 A and B at 18~cm.  The contours are plotted
at intervals of -0.15, 0.15, 0.3, 0.6, 1.2, 2.4, 4.8, 9.6 mJy/beam.
The 4~mas restoring beam is shown in the lower left.  }
\end{figure}

\newpage

{\it This page contains additional material from the poster which did
not fit in the published proceedings.}

\begin{center}
{\bf Simulation: The Potential of Lensed Time Delays for Cosmology}
\end{center}

\begin{itemize}
\item The fundamental cosmological result from lensed time delays
is not $H_0$, but the angular diameter distance to the lens 
(Narayan 1991). 

\item If we could measure the angular diameter distance 
to several lenses, then we could set limits on the matter density of
the universe $\Omega_m$ and the cosmological constant
$\Omega_\lambda$.

\item  Simulation details: Assuming $\Omega_m=0.3$, 
$\Omega_\lambda=0.7$, use the redshift for each lens to calculate its
angular diameter distance.  Add 15\% Gaussian errors to each lens
distance.  Then use distances to fit for $\Omega_m$ and
$\Omega_\lambda$.  The figure shows the resulting 68\% confidence
intervals.

\item As of early July 1999, there were 15 lenses with point images
and known lens and source redshifts.  If a time delay could be
measured for each of these, we could have comparable results to 42
Type Ia Supernovae (Perlmutter et al. 1999).

\end{itemize}

\begin{figure}
\plotfiddle{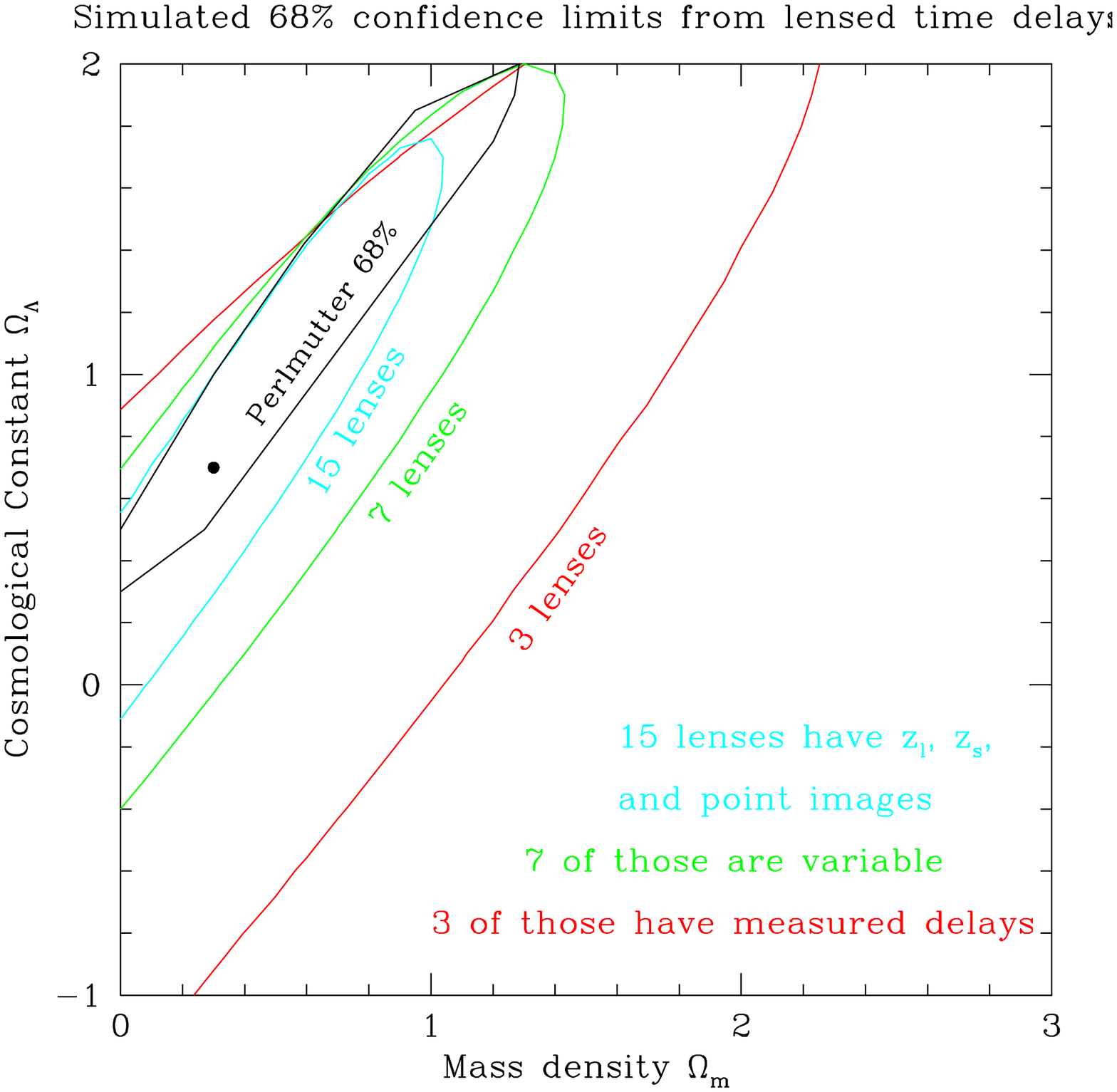}{3in}{0}{40}{40}{-110}{-40}
\end{figure}

\end{document}